\documentclass[letterpaper,10pt,twocolumn,twoside]{article}

\usepackage{cite}
\usepackage{amsmath}
\usepackage{graphicx}
\usepackage[latin1]{inputenc}
\usepackage{amsfonts}
\usepackage{amssymb}
\usepackage[usenames]{color}

\usepackage{fancyhdr}
\usepackage{sectsty}
\usepackage[small,bf,sf,up]{caption}

\allsectionsfont{\sffamily}
\allsectionsfont{\normalsize\bfseries}
\begin{document}


\title{BioVEC: A program for Biomolecule Visualization with Ellipsoidal Coarse-graining}

\author{Erik Abrahamsson$\thanks{E-mail: erik@phas.ubc.ca}$, Steven S. Plotkin$\thanks{E-mail: steve@phas.ubc.ca}$, \\
\\
Department of Physics and Astronomy, \\
The University of British Columbia, \\
6224 Agricultural Road, Vancouver, B.C. V6T 1Z1, Canada \\
\\
{\footnotesize Accepted for publication in Journal of Molecular Graphics and Modelling} \\
{\footnotesize doi:10.1016/j.jmgm.2009.05.001}}

\date{May 18, 2009}
\maketitle

%

\begin{minipage}[c]{0.90\textwidth}
\sloppy Biomolecule Visualization with Ellipsoidal Coarse-graining (BioVEC) is a tool for visualizing molecular dynamics simulation data while allowing coarse-grained residues to be rendered as ellipsoids. BioVEC reads in configuration files, which may be output from molecular dynamics simulations that include orientation output in either quaternion or ANISOU format, and can render frames of the trajectory in several common image formats for subsequent concatenation into a movie file. The BioVEC program is written in C++, uses the OpenGL API for rendering, and is open source. It is lightweight, allows for user-defined settings for  and texture, and runs on either Windows or Linux platforms.
\end{minipage}




\section{Introduction}
\label{intro}

The use of coarse-graining methods in biomolecular simulation is progressively increasing~\cite{Toz05:144}, mainly to circumvent the current time-scale problem that exists for all-atom molecular dynamics simulations, and to allow for the computational modeling of ever larger systems which address biologically relevant questions. However as the coarse-grained constituents encompass more atoms, the resulting residual objects become poorly approximated by spheres with corresponding isotropic potentials, while the steric character of the residues becomes more important. Such a scenario would exist for the coarse-graining of bases in DNA, or the coarse-graining of helices in a protein.

For these reasons, interest in ellipsoidal coarse-graining, perhaps the simplest non-spherical generalization, has recently increased~\cite{Gay81:3316,Ber98:8,Eve03:041710,Par05:194111,Bab06:174708,Pli95:1,Ber08:463101,Ste95:3098}. Interest in industrial applications such as the manufacture of nanoparticle films has also motivated studies of ellipsoidal coarse-graining~\cite{Gre08:1,BrownWM09}.

The molecular dynamics (MD) of rigid objects raises some new practical challenges, including a 
\pagebreak
\\
\\
\\
\\
\\
\\
\\
\\
\\
\\
singularity-free representation of rigid body rotation~\cite{Eva77:317,Eva77:327,Nit00:83,Cou04:1849} and corresponding energy conserving scheme for symplectic integration~\cite{Mil02:8649,Shi04:124,Kam05:224114}. This is best done using quaternions\cite{Ham44:424,Kar07:595}, which MD simulation programs such as LAMMPS~\cite{Pli95:1} currently allow one to employ. 

Quite apart from the issues of coarse-graining and molecular dynamics, visualization tools for rendering thermal ellipsoids representing Debye-Waller factors have enjoyed a long and successful tradition\cite{Bur96:6895}. However, in spite of widely used standards in crystallography, as well as the above-mentioned progress in coarse-graining and molecular dynamics of rigid bodies, to the authors' knowledge no currently available program allows for the straightforward visualization of  the dynamical evolution of coarse-grained systems with ellipsoidal constituents. To facilitate our own analysis of such coarse-grained models, we have decided to develop our own program. The program is portable to either Windows or Linux systems, and could in principle be incorporated into many of the more versatile and widely-used visualization programs such as VMD~\cite{Hum96:33}, Molmol~\cite{Kor95:51}, Molscript~\cite{Kra91:946}, PyMOL~\cite{pymol}, Raster3D\cite{Mer97:505}, Jmol\cite{jmol}, and Rasmol\cite{rasmol}.

\section{Methods}
\label{methods}

The program BioVEC presented in this paper, is a program to visualize the results of MD simulations of biomolecules that have been coarse-grained with ellipsoids. The functionality of BioVEC is simple and basic, as its purpose is not to compete with current molecular visualization packages but specifically to allow the scientist to visualize dynamical processes involving ellipsoidal coarse-graining. BioVEC automatically creates graphical images, which can be compiled to movie files in an external editor, such as Slide Show Movie Maker\cite{ssmm} (Windows) or MJPEG Tools\cite{mjpeg} (*NIX). It is also possible for the user to manually step through the simulation, and to save the current view of the molecule to an image file in several formats, including  {\tt png}, {\tt bmp}, and {\tt jpg}. The representation used in BioVEC differs from techniques such as cartoon and ribbon representation, as these techniques aim to smooth out details, in all-atom simulations for example. The aim of BioVEC is rather to show all the features of already coarse-grained simulation data.

BioVEC reads in molecular dynamics simulation data from a file containing the positions of the coarse-grained species and the orientations of the ellipsoids for all time steps in the molecular dynamics simulation. The input format for BioVEC is based on the output dump file format in LAMMPS\cite{Pli95:1}, which is of the form:

{\footnotesize \begin{quote}{\ttfamily species-no \hspace{0.5em} species-type \hspace{0.5em} position~$\times$~3 \hspace{0.5em} orientation~$\times$~4}\end{quote}}

for each time step. The program also reads in a file with information on the topology of the molecule, in order for the program to display bonds between the species. In the current version of the program, covalent bonds are not allowed to break or form between different time steps. The information in the topology file comes from the LAMMPS input file, and is of the form 

{\footnotesize \begin{quote}{\ttfamily bond-no  bond-type  1st-species 2nd-species 1st-origin 2nd-orgin}\end{quote}}

\sloppy where the bond goes from the {\ttfamily 1st-origin} on the {\ttfamily 1st-species} to the {\ttfamily 2nd-origin} on the {\ttfamily 2nd-species}, corresponding to the {\ttfamily species-no} in the simulation data file. Each ellipsoid can have an arbitrary number of bond origins, as specified in the input file. The {\ttfamily 1st-origin} and {\ttfamily 2nd-origin} entries in the topology file are optional if the species only have one bond origin each. 

The user supplies the names of the simulation data and topology files in an input file with information about the coarse-graining of the system. This file should contain the number of coarse-grained species, the radii of the different ellipsoids and spheres, as well as the vector from the centre of mass of the ellipsoid to the location of the bond connecting the ellipsoidal residue to its covalently-bonded partner. Spheres are treated as ellipsoids with all three radii equal. Information about the colouring and texture of the species is also entered into the input file.  

\sloppy The BioVEC program can also read and visualize ANISOU anisotropic temperature factor records in PDB standard form\cite{anisou}.

\subsection{Quaternions}

Quaternions\cite{Ham44:424} are frequently used in computer graphics to represent rotations and orientation in 3-dimensional space. One of the main advantages of quaternions over other representations, such as Euler angles, is the absence of singularities in the quaternion representation. 

A quaternion is a four-dimensional extension to the complex numbers,
\begin{equation}
	{\bf q} = q_0{\bf u} + q_1{\bf i} + q_2{\bf j} + q_3{\bf k}.
	\label{eq:quat}
\end{equation}
where $q_0$, $q_1$, $q_2$, and $q_3$ are real numbers, and {\bf u}, {\bf i}, {\bf j}, and {\bf k}
are such that ${\bf i}^2 = {\bf j}\,^2 = {\bf k}^2 = {\bf ij\:k} =
-{\bf u}^2 = -1$, and ${\bf ij} = -{\bf j\,i} = {\bf k}$. Quaternions are often written in a more compact vector form,  
\begin{equation}
	{\bf q} = [q_0, q_1, q_2, q_3],
	\label{eq:vecquat}
\end{equation}\\
where the quantities $q_0$, $q_1$, $q_2$, and $q_3$ again are real numbers. 

The relation between the quaternion representation and the rotation and angle $\alpha$ about a unit axis, ${\bf \hat{n}} = x {\bf\hat{x}} + y {\bf\hat{y}} + z {\bf\hat{z}}$ is given by (see for example~\cite{Wolfram})  
\begin{gather}
	s = {(q_1^2 + q_2^2 + q_3^2)}^{1/2}, \notag \\ 
	x = q_1/s, \notag \\
  y = q_2/s, \\
  z = q_3/s, \notag \\ 
    \alpha = 2\cos^{-1}(q_0) \: , \notag
  \label{eq:rot}
\end{gather}
for $s \neq 0$. Note that for the $q_i$ to represent a rotation $\sum_{i=0}^3 q_i^2 = 1$. The orientation of the different ellipsoids are read by BioVEC from the simulation dump-file in vector form (Eq.~\ref{eq:vecquat}), together with the position and ellipsoid type.

\subsection{Bond vectors}

\begin{figure}
	\centering
		\includegraphics[width=0.45\textwidth]{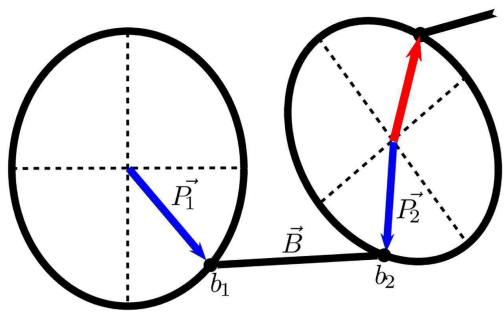} 
	\caption{Representation of the bond vectors. The bond $\vec{B}$ is formed between the points $b_1$ and $b_2$, given by the bond vectors $\vec{P_1}$ and $\vec{P_2}$. A second bond vector for the right ellipsoid is indicated by a red arrow. The bond vectors are given relative to the principal axis (dotted lines) of the ellipsoids.}
	\label{fig:bond}
\end{figure}

The bonds between ellipsoidal species are drawn between the same points on the ellipsoids as were used in the MD simulation. The vectors from the centre of mass of the different types of ellipsoids to the location where the covalent bond joins the ellipsoid (Fig.~\ref{fig:bond}), are supplied by the user in the configuration file. The vectors are given in the principle (body-centered) basis of the ellipsoid for each bond origin on each ellipsoid species. 

\section{Implementation}
\label{impl}

The BioVEC program is written in C++, using the OpenGL API for rendering. The GLUT library is used for creating menus. The user input is command line driven, and the interaction with the program is via keyboard commands and mouse menus.

Graphical images from the program can be saved to file in several different formats through the DevIL library\cite{devil}. Images can be saved automatically for each time step of the simulation as the program reads through the data file, or the current view of the molecule can be saved by the user.

\sloppy The BioVEC source can be compiled and run on both the Windows and Linux platforms.

\section{Results and Discussion}
\label{disc}

\begin{figure}
	\centering
		\includegraphics[width=0.45\textwidth]{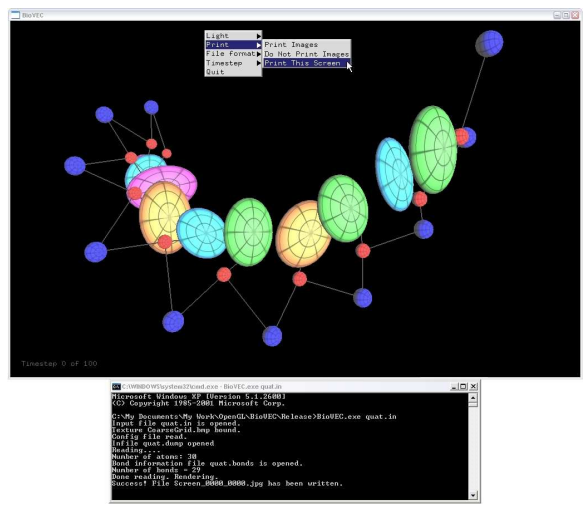} 
	\caption{View of the BioVEC interface, running on Windows XP.}
	\label{fig:gui}
\end{figure}

Figure~\ref{fig:gui} shows an example of BioVEC running under Windows XP, depicting the console window, the OpenGL window, and the GLUT mouse menu. The molecule can be rotated and zoomed with either the keyboard or the mouse, and the time stepping can be done automatically or one step at at time by the user. The user can let the program automatically print the current view to an image file, or can choose to manually print the current view with a press of a key. This way stereo views of the molecule can be created. 

\begin{figure}
	\centering
		\includegraphics[width=0.45\textwidth]{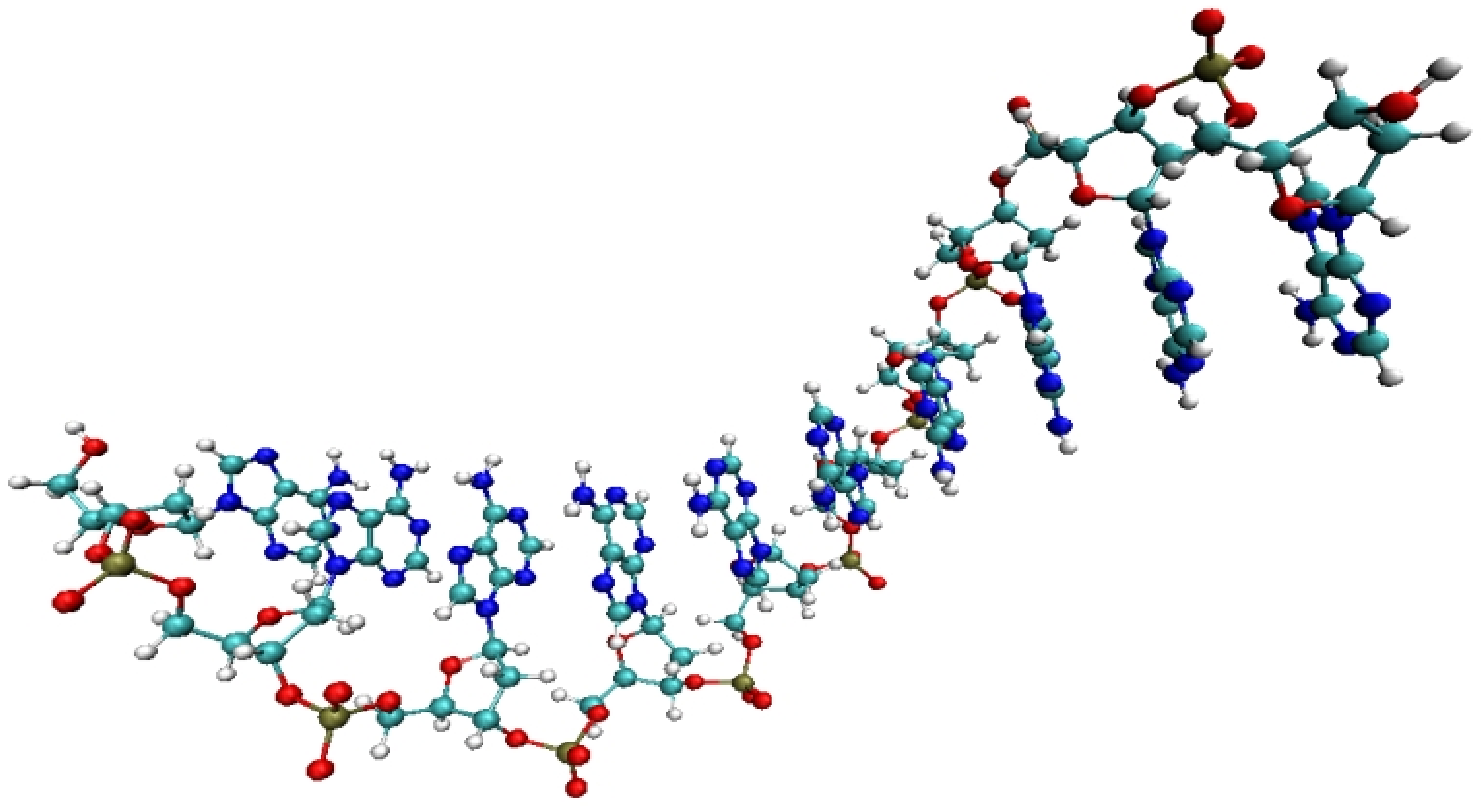} 
		\includegraphics[width=0.45\textwidth]{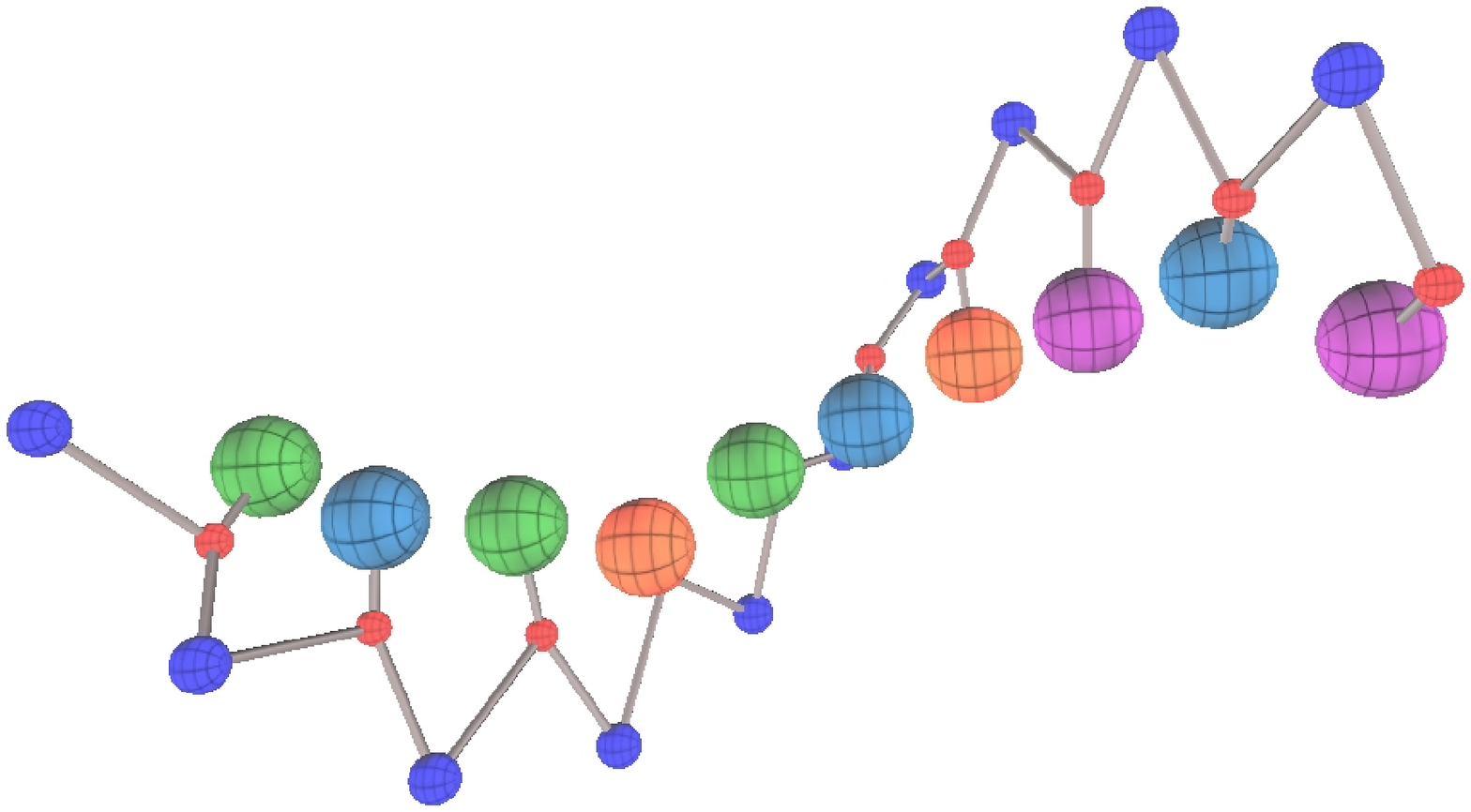} 
		\includegraphics[width=0.45\textwidth]{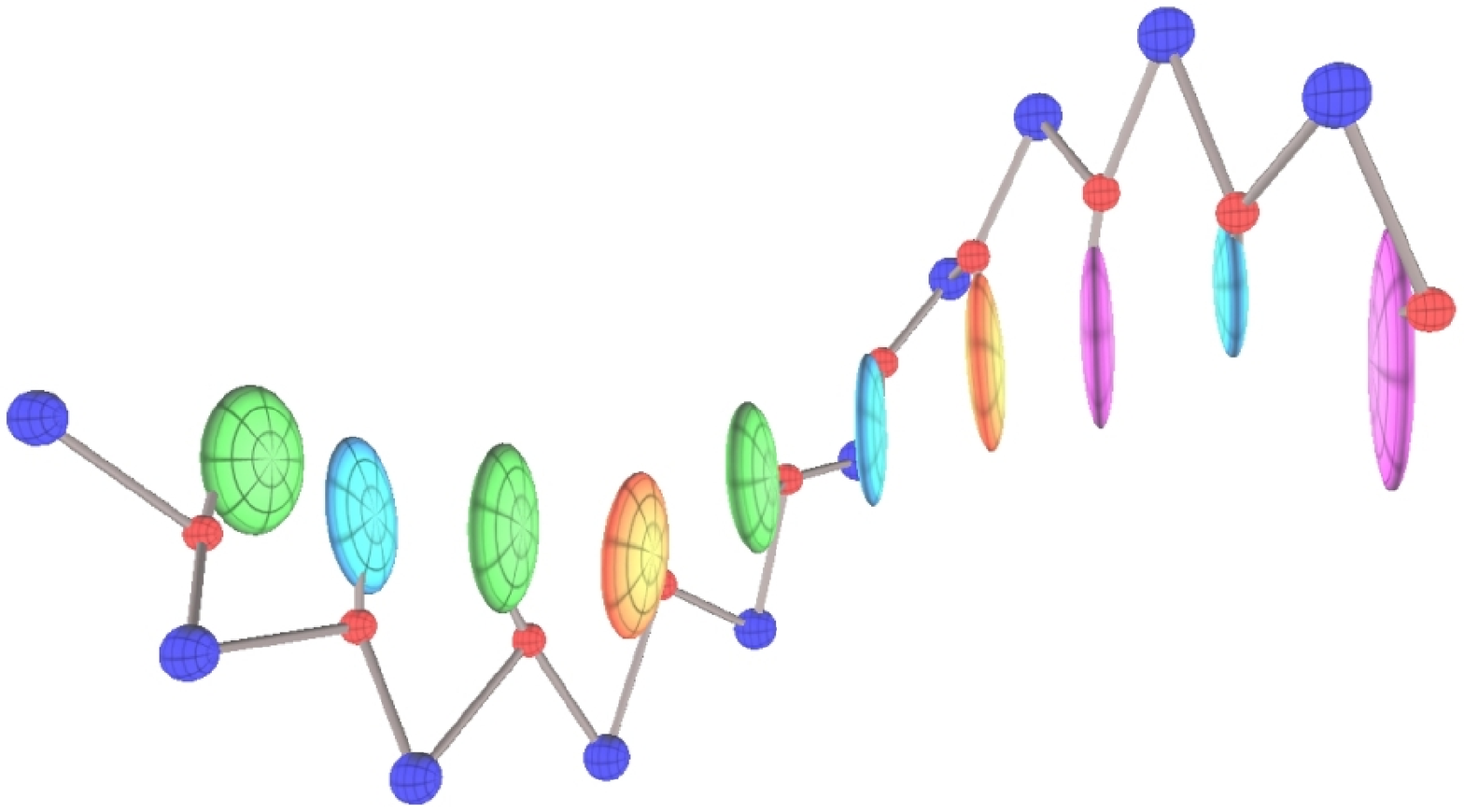} 
	\caption{Comparison of three different visualization models for single-stranded DNA. All-atom CPK representation rendered in VMD; coarse-grained ball-and-stick model; and coarse-grained ellipsoidal model, both rendered in BioVEC.}
	\label{fig:ssdna}
\end{figure}

The advantage of using a program, such as BioVEC, capable of visualizing ellipsoids, is seen in Fig.~\ref{fig:ssdna}, where the all-atom model of single-stranded DNA is shown alongside with a coarse-grained ball-and stick model, and an ellipsoidal model, both visualized in BioVEC. In the all-atom model the orientation of the bases can be seen, while this information is lost when displaying the coarse-grained DNA strand with the ball-and stick model. The ellipsoidal visualization model used in BioVEC, however, retains the orientational information of the all-atom representation.

\begin{figure}
	\centering
		\includegraphics[width=0.45\textwidth]{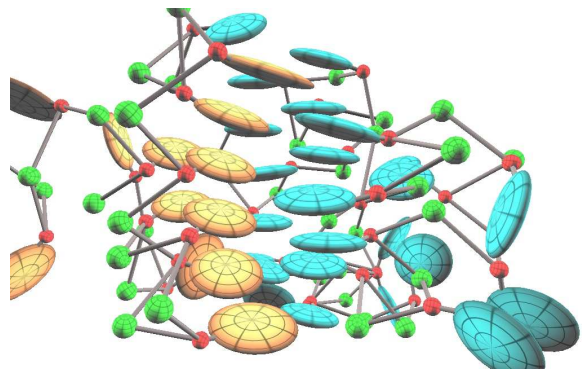} 
	\caption{BioVEC visualization of single stranded DNA consisting of adenine and thymine, forming a hairpin.}
	\label{fig:dsdna}
\end{figure}

Figure~\ref{fig:dsdna} shows a collapsed globule of single-stranded DNA with block-copolymeric sequence A($\times 15$)- T($\times 15$). Information that would not have been seen with a coarse-grained ball-and-stick rendering, such as partial hair-pin formation along with orientation of the stacked base pairs, is clearly visualized in BioVEC. 

\begin{figure}
	\centering
		\includegraphics[width=0.45\textwidth]{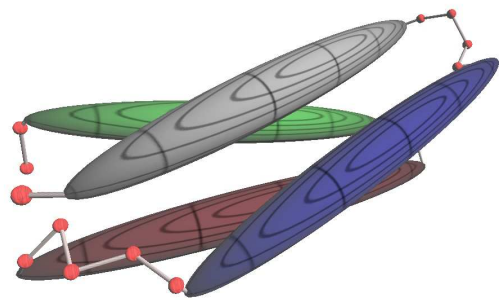} 
	\caption{BioVEC visualization of the homodimeric ROP protein 1YO7 four helix bundle, with the helices represented by ellipsoids. Multiple bond origins are present for this system, and represent the points where the backbone chain enters and exits the helices.}
	\label{fig:1yo7}
\end{figure}

The input format of the program also allows for multiple bond origins, as shown in Fig.~\ref{fig:1yo7}, where a 4-helix bundle is coarse-grained so that the helices are modeled by effective prolate ellipsoids. Investigating the folding mechanism of such a coarse-grained system allows one to investigate the assumptions of the diffusion-collision model\cite{Kar94:650}, for example. Renderings with multiple bond origins may be useful for characterizing the constraints due to intermolecular chemical bonds between biomolecules.

\begin{figure}
	\centering
		\includegraphics[width=0.45\textwidth]{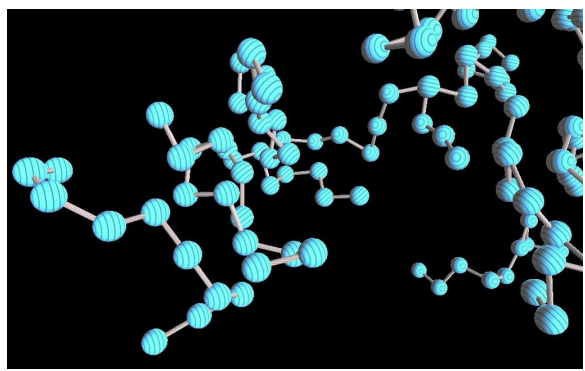} 
	\caption{BioVEC visualization of polymer glass\cite{War09}. The simulation contains 100 polymers with ten atoms in each polymer. Each polymer is visualized with different orientation of the texture to simplify the interpretation of the image.} 
	\label{fig:betaHP}
\end{figure}

Separate strands of coarse-grained polymers can be visualized in BioVEC. An example of this is shown in Fig.~\ref{fig:betaHP}, where a polymer glass melt, consisting of 100 polymer strands with ten atoms in each strand\cite{War09}, is visualized. By giving all the atoms on a single polymer the same orientation, the separate polymers in the glass can be identified. The information about the bonds, and thereby the separate polymers, is given in the topology file.

\begin{figure}
	\centering
		\includegraphics[width=0.45\textwidth]{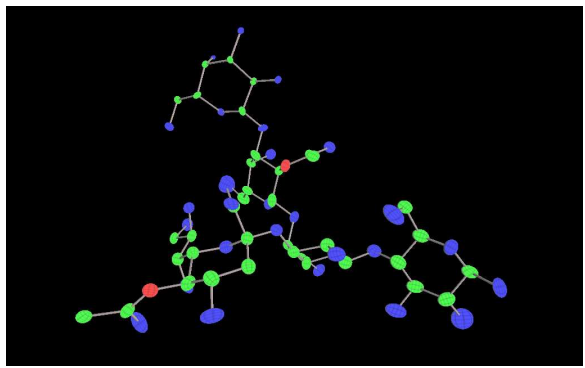} 
	\caption{BioVEC visualization of the thermal anisotropy of GM1-pentasaccharide from a complex with the cholera toxin B-pentamer\cite{Mer97:505}.}
	\label{fig:sugarG}
\end{figure}

BioVEC can also read and visualize PDB files with ANISOU data. Figure~\ref{fig:sugarG} shows the thermal ellipsoid representation of GM1-pentasaccharide from a complex with the cholera toxin B-pentamer refined with anisotropic displacement parameters at 1.25~\AA{} resolution (PDB data taken from a Raster3D\cite{Mer97:505} example file).

\begin{figure}
	\centering
		\includegraphics[width=0.45\textwidth]{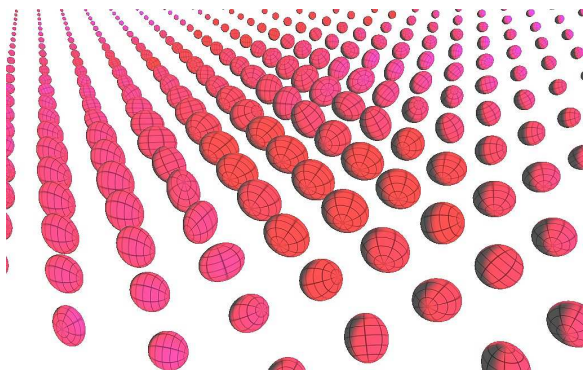} 
	\caption{BioVEC visualization of the dielectric tensors at a cross-section of ubiquitin.\cite{Gue09}. The principal axis of the ellipsoids are given by the inverse of the eigenvalues of the dielectric tensor.}
	\label{fig:dielec}
\end{figure}

BioVEC can be used to visualize ellipsoids for a wide range of applications. Figure~\ref{fig:dielec} shows the orientation of the dielectric tensors at a cross-section through the centre of ubiquitin\cite{Gue09}. The ellipsoids represents the local dielectric constant, with the orientation given by treating the dielectric tensor as a rotation matrix. The sizes of the principal axis are given by the reciprocals of the eigenvalues, to emphasize the low dielectric constant in the centre of the protein.

\section{Conclusion}
\label{conc}

The BioVEC program presented here is a simple but useful tool for visualizing bio-molecules that are coarse-grained with ellipsoids. There are many program packages available for molecular visualization, but none of these can, to our knowledge, display biomolecules coarse-grained with ellipsoids. Some programs, such as Raster3D\cite{Mer97:505} and PyMOL\cite{pymol}, allow for the visualization of anisotropic temperature factors in the form of ellipsoids. However, they are currently not suitable for straight-forward visualization of the dynamics of coarse-grained ellipsoidal systems with the orientation represented by quaternions. The BioVEC program therefore focuses on this feature, making it a small and user-friendly program for visualization of coarse-grained molecular dynamics data. 

Future revisions of BioVEC should include a graphical user interface (GUI), in order to improve both the functionality and the ease-of-use. Improvement and optimization of the rendering algorithm should accelerate rendering of large molecules, or long simulation runs.

BioVEC is open source, and is distributed under the GNU GPL licence. The source code and Windows executable are available for download at http://www.phas.ubc.ca/$\sim$steve/BioVEC/.

\section{Acknowledgments}
\label{ack}

E.A gratefully acknowledges support from the Sweden-America Foundation and the Olle Engkvist Foundation. S.S.P. gratefully acknowledges support from the Natural Sciences and Engineering Research Council and the A. P. Sloan Foundation. We are grateful to Alex Morriss-Andrews, Will Guest, and Mya Warren for supplying simulation data and for helpful discussions.

\bibliographystyle{elsart-num}

\end{document}